\documentclass[%
 reprint,
superscriptaddress,
 amsmath,amssymb,
 aps, prb
]{revtex4-2}

\usepackage{graphicx}% Include figure files
\usepackage{dcolumn}% Align table columns on decimal point
\usepackage{bm}% bold math
\usepackage{braket}
\usepackage[colorlinks=true, allcolors=blue]{hyperref}
\usepackage{cleveref}
\begin{document}
\bibliographystyle{apsrev4-2.bst}
\preprint{APS/123-QED}

\title{Defect Phonon Renormalization during Nonradiative Multiphonon Transitions in Semiconductors}
\author{Junjie Zhou}
\thanks{These authors contributed equally.}
\affiliation{School of Microelectronics, Fudan University, Shanghai 200433, China}
\author{Shanshan Wang}
\thanks{These authors contributed equally.}
\affiliation{Key Laboratory of Computational Physical Sciences (MOE), Fudan University, Shanghai 200433, China}
\author{Menglin Huang}
\email{menglinhuang@fudan.edu.cn}
\affiliation{School of Microelectronics, Fudan University, Shanghai 200433, China}
\affiliation{Key Laboratory of Computational Physical Sciences (MOE), Fudan University, Shanghai 200433, China}
\author{Xin-Gao Gong}
\affiliation{Key Laboratory of Computational Physical Sciences (MOE), Fudan University, Shanghai 200433, China}
\author{Shiyou Chen}
\email{chensy@fudan.edu.cn}
\affiliation{School of Microelectronics, Fudan University, Shanghai 200433, China}
\affiliation{Key Laboratory of Computational Physical Sciences (MOE), Fudan University, Shanghai 200433, China}

\begin{abstract}
As a typical nonradiative multiphonon transition in semiconductors, carrier capture at defects is critical to the performance of semiconductor devices. Its transition rate is usually calculated using the equal-mode approximation, which assumes that phonon modes and frequencies remain unchanged before and after the transition. Using the carbon substitutional defect ($\text{C}_\text{N}$) in GaN as a benchmark, here we demonstrate that the phonon renormalization can be significant during defect relaxation, which causes errors as large as orders of magnitude in the approximation. To address this issue, we consider (i) Duschinsky matrix connecting the initial-state and final-state phonons, which accounts for the changes in phonon modes and frequencies; and (ii) the off-diagonal contributions in total transition matrix element, which incorporates the cross terms of electron-phonon interactions between different modes. With this improvement, the calculated transition rates show agreements with experimental results within an order of magnitude. We believe the present method makes one step forward for the accurate calculation of multiphonon transition rate, especially in cases with large defect relaxations.
\end{abstract}

%\keywords{Suggested keywords}%Use showkeys class option if keyword
                              %display desired
\maketitle

%\tableofcontents

\par \textit{Introduction}\textemdash Carrier capture at defects is a nonradiative multiphonon transition of carriers from band edges to defect states, which is recognized as a critical process influencing the performance of semiconductor devices \cite{defectHans1998}. \textit{For instance}, continuous hole capture at oxide defects in p-type field-effect transistors causes a negative shift in the threshold voltage, known as a typical reliability issue \cite{GRASSER201239}. Therefore, an accurate calculation of defect-assisted carrier capture rate has attracted wide interests from the researchers in solid-state and semiconductor physics communities \cite{henry1977nonradiative, zhang2021minimizing,DemchenkoPRL}.

\par According to the theory pioneered by Huang and Rhys in 1950 \cite{huang1950theory,kubo1955application,passler1974description,passler1975description}, the transition is simultaneously accompanied by a change (lattice relaxation \cite{huang1981lattice}) in the equilibrium structures of the defect system. To calculate the transition rate, the Huang-Rhys theory requires a specific structure for expanding the Hamiltonian of electron-phonon interaction \cite{huang1981lattice}. Usually, either the initial-state structure before the transition or the final-state structure after the transition is chosen in the calculation. This is currently known as equal-mode approximation based on the initial-state or final-state phonons, i.e., either the initial-state phonons or the final-state phonons are considered \cite{Alkauskas2021}, and the phonon renormalization during the transition is neglected. With this approximation, significant progresses have been made in the first-principles calculation methods of defect-assisted nonradiative multiphonon transition rate \cite{Shi2012, shi2015comparative, xiao2020anharmonic,qiuchen}. On the other hand, equal-mode approximation is also used for calculating the radiative luminescence lineshape of color center quantum defects \cite{alkauskas2014NV,Giulia2021,PhysRevB.110.075303,geoffory-appro}.

\par Despite the progress, the accuracy of equal-mode approximation has not been clearly assessed yet. The precondition of the approximation is that the lattice relaxation is small, so that the phonon renormalization can be negligible \cite{RevModPhys.31.956,Galli1}, but defects in crystals can experience substantial structural relaxations upon charge state changes \cite{whalley2021giant,zhang2022origin}, which may affect phonons significantly. To date, an effective first-principles method considering the phonon renormalization in defect-induced multiphonon transitions is still missing, leaving the equal-mode approximation as a compromise in the current stage \cite{Alkauskas2021}. If defect relaxations are non-negligible, a general method beyond the equal-mode approximation is urgently required for calculating transition rates accurately.

\par In this Letter, we revisit the formalism of equal-mode approximation in the Huang-Rhys theory and quantitatively evaluate the impact of defect phonon renormalization using first-principles calculations. Our results show the widely-adopted equal-mode approximation can cause errors by orders of magnitude in nonradiative transition rates and thus is not proper for describing the multiphonon processes in semiconductors. To solve this issue, we incorporate the Duschinsky matrix between initial-state and final-state phonons, and meanwhile consider the off-diagonal contributions to transition matrix element. As a result, this approach reduces the error significantly, and the calculated capture coefficient shows agreement with experimental values within an order of magnitude.

\par \textit{Assessing the validity of equal-mode approximation}\textemdash According to Fermi's golden rule, nonradiative multiphonon transition rate can be calculated by \cite{huang1981lattice},
\begin{equation}\label{fermi_golden_rule}
    r=\frac{2\pi}{\hbar}\rho(E_{im}) \left| \left< \Phi _{im}\left| H_{eL} \right|\Phi _{fn} \right> \right|^2\delta (E_{im}-E_{fn}),
\end{equation}
where $H_{eL}$ is the Hamiltonian of electron-phonon interaction, $\rho(E_{im})$ is the initial state phonon occupation. The transition matrix element between initial and final state within the static coupling scheme can be written as \cite{passler1974description,passler1975description},
\begin{equation}\label{static_coupling}
    \begin{aligned}
        &\left< \Phi _{im}\left| H_{eL} \right|\Phi _{fn} \right> \\
        &=\int{\mathrm{d}}x\int{\mathrm{d}}Q\,\,\chi _{m}^{*}(\bar{Q})\varphi _{i}^{*}(x)H_{eL}(x,Q)\varphi _f(x)\chi _n(\bar{\bar{Q}}),
    \end{aligned}
\end{equation}
where $\chi\left(Q\right)$ is the vibrational wavefunction, and $\varphi\left(x\right)$ is the electronic wavefunction at a given configuration $Q$. $i$ and $f$ are the initial- and final-state indices for the electronic states, and $m$ and $n$ are those for the vibrational states. There are two sets of Q-space: the configurations $\bar{Q}$ based on the structure before the transition and that $\bar{\bar{Q}}$ after the transition. Considering the Q-independent term $\varphi\left(x\right)$ and the Q-dependent term $\chi\left(Q\right)$ due to static coupling, it is useful to adopt the so-called linear coupling approximation \cite{passler1974description}, which expands the Hamitonian $H_{eL}\left(x,Q\right)$ linearly with $Q$ in mode $k$,
\begin{equation}\label{linear_approx}
    H_{eL}(x,Q)=\sum_k{\frac{\partial H_{eL}}{\partial Q_k}}Q_k,
\end{equation}
where the coefficient $\partial H^{k}_{eL} / \partial Q_{k}$ is independent of $Q$. Then Eq. (\ref{static_coupling}) becomes,
\begin{equation}\label{separation}
    \begin{aligned}
        &\left< \Phi_{im}  \left| H_{e L} \right| \Phi_{fn}\right> = \\
        &\sum_k \int d x \varphi_i^* (x) \frac{\partial H_{e L}}{\partial Q_k} \varphi_f(x) \int d Q \chi_m^*(\bar{Q}) Q_k \chi_n(\bar{\bar{Q}})  
    \end{aligned}
\end{equation}
and this is a separation of the total transition matrix element into an electron-phonon coupling term $W_{k}^{\{ if \} }=\left< \varphi_i \left| \partial H_{eL}/\partial Q_k \right| \varphi_f \right> $ and a term $\left<\chi_m\left| Q_k \right| \chi_n \right>$ associated with lattice relaxation. 

\par Due to lattice relaxation, the equilibrium structure $\bar{Q}_0$ before the transition and that $\bar{\bar{Q}}_0$ after the transition are different, so their corresponding phonon wavefunctions differ as well. This leads to a highly multi-dimensional integral in Eq. (\ref{separation}). Equal-mode approximation is then used to reduce the computational cost and facilitate first-principles calculations \cite{Shi2012,PhysRevB.110.075303,Alkauskas2021}. In practice, there are two typical choices when this approximation is adopted. (\textbf{i}) The equilibrium defect structure $\bar{Q}_0$ before the transition is set as the coordinate origin in Eq. (\ref{linear_approx}), then all phonon modes and wavefunctions are calculated based on this initial-state structure. Therefore, both $\chi_m$ and $\chi_n$ are eigenstates as functions of $\bar{Q}$, leading to the lattice relaxation matrix expressed as $\left< \chi_m(\bar{Q})\left| \bar{Q} \right| \chi_n (\bar{Q}) \right>$. This is referred as equal-mode approximation based on initial state. (\textbf{ii}) Alternatively, the equilibrium defect structure $\bar{\bar{Q}}_0$ after the transition is set as the origin, with all phonon modes and wavefunctions calculated based on the final-state structure. Here, the lattice relaxation matrix takes the form $\left< \chi_m(\bar{\bar{Q}})\left| \bar{\bar{Q}} \right| \chi_n (\bar{\bar{Q}}) \right>$. This is referred as equal-mode approximation based on final state.

\par If equal-mode approximation is valid, defect relaxation should affect phonon frequencies minorly and the results calculated by equal-mode approximation based on initial state and final state should be similar. Here we will use carbon substitutional defect ($\text{C}_{\text{N}}$) in GaN as a benchmark for testing the validity of the approximation, \textit{i.e.}, we will compare the hole capture coefficients calculated by the two choices of equal-mode approximation, respectively. The first-principles calculations are performed under the formalism of projector-augmented wave (PAW) implemented in  Vienna \textit{ab initio} simulation package (\textsc{vasp}) \cite{kresse1996efficiency,kresse1999ultrasoft}. Hybrid functional of Heyd, Scuseria, and Ernzerhof (HSE) \cite{heyd2003hybrid,10.1063/1.2204597} is adopted with the exchange parameter set to be 29\%. More computational details and the formalism can be found in Supplemental Material \cite{supplement}. \nocite{1940264,zhanghaishan2017,Freysoldt2014,FNV,Huang_2022}

\begin{figure}[htbp]
	\centering
	\includegraphics[width=0.48\textwidth]{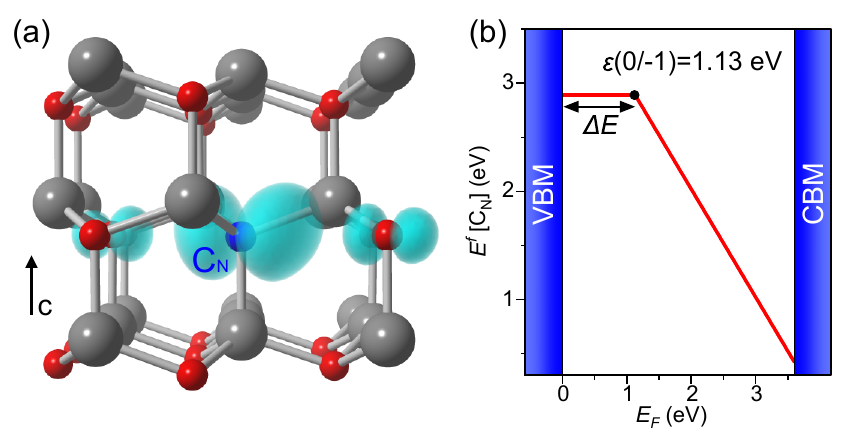}
	\caption{(a) Atomic structure of neutral $\text{C}_{\text{N}}$ in GaN. The contour plot shows the norm-squared wavefunction of the unoccupied (spin-down) impurity state. (b) Formation energy of $\text{C}_{\text{N}}$ as a function of Fermi level. }
	\label{fig1}
\end{figure}

\begin{figure*}[htbp]
	\centering
	\includegraphics[width=0.65\textwidth]{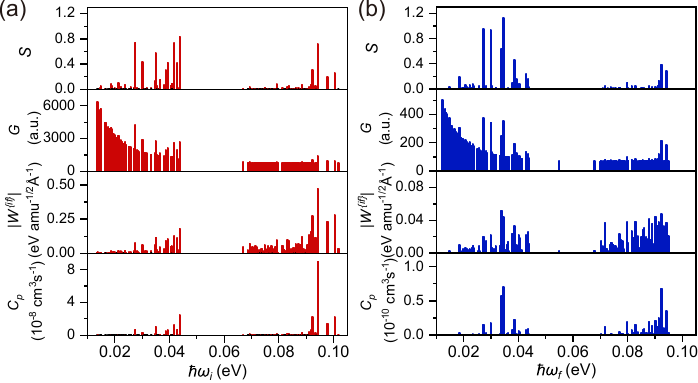}
	\caption{The contribution of different phonon modes to the Huang-Rhys factor ($S$), lineshape function ($G$), electron-phonon coupling matrix element ($W^{\{ if \} }$), and hole capture coefficient ($C_p$) when equal-mode approximation based on (a) initial state and (b) final-state is adopted. When equal-mode approximation based on initial state is adopted, all the phonon modes and frequencies are calculated based on the structure of $\text{C}_{\text{N}}^{-1}$; when equal-mode approximation based on final state is adopted, those are calculated based on the structure of $\text{C}_{\text{N}}^{0}$.}
	\label{fig2}
\end{figure*}

\par The equilibrium structure of neutral $\text{C}_{\text{N}}$ is shown in Fig. \ref{fig1}(a), where the blue atom shows the center of $\text{C}_{\text{N}}$. In neutral state, $\text{C}_{\text{N}}$ defect states are half occupied, \textit{i.e.}, the occupied state is at the spin-up channel and the unoccupied state is at the spin-down channel. When $\text{C}_{\text{N}}$ becomes $-1$ charged, the spin-down state also becomes occupied. Therefore, the unoccupied state in neutral $\text{C}_{\text{N}}$ is associated with the $(0/-1)$ transition, with the norm-squared wavefunction of this state depicted in Fig. \ref{fig1}(a). The cyan contour indicates that this defect state is localized and consists mainly of the C 2\textit{p} orbitals. For neutral $\text{C}_{\text{N}}$, the four nearest-neighbor Ga atoms move inward by 1\% in average compared to the Ga-N bond length in equilibrium GaN bulk. For $-1$ charged $\text{C}_{\text{N}}$, two of the four C-Ga bonds are extended by 0.093 Å. The change in C-Ga bond is the main contribution to the lattice relaxation when defect charge state is switched between 0 and $-1$. Fig. \ref{fig1}(b) shows $\text{C}_{\text{N}}$ produces a $(0/-1)$ transition level at 1.13 eV above the valence band maximum (VBM), which is similar to other theoretical results \cite{alkauskas2014first,turiansky2021nonrad,demchenko2020koopmans} but slightly larger than the experimental results \cite{reshchikov2018two,narita2018origin}. When multiphonon transition occurs at $\text{C}_{\text{N}}$ defect, a hole is captured from the VBM level to the $(0/-1)$ level, which changes the defect charge state from $-1$ to 0 meanwhile induces the aforementioned lattice relaxation.

\par In the equal-mode approximation based on initial state, both the Huang-Rhys factor ($S_k=\omega _k\Delta Q_{k}^{2}/2\hbar $) and lineshape function (Eq. \ref{lineshape}) are derived from the phonons calculated at the equilibrium structure of $-1$ charged $\text{C}_{\text{N}}$. The lineshape function in multiphonon transition can be defined as \cite{turiansky2021nonrad},
\begin{equation}\label{lineshape}
    G\left( \Delta E \right) =\sum_m{\rho _{m}}\sum_n{\left| \left< \chi _m\left| Q \right|\chi _n \right> \right|^2}\delta (\Delta E+{E_{m}}-{E_{n}}),
\end{equation}
where $\Delta E$ is the electronic transition energy, $E_m$ and $E_n$ are the $m^{th}$ phonon energy of the initial state and $n^{th}$ phonon energy of the final state. Comparing the calculated phonon density of states (DOS) of $\text{C}_{\text{N}}^{-1}$ and GaN bulk shown in Fig. S1, we find $\text{C}_{\text{N}}^{-1}$ defect system introduces high-frequency localized modes around $100 \text{ meV}$. However, these modes do not coincide exactly with the direction of the lattice relaxation. In fact, a variety of phonon modes contribute to the overall lattice relaxation, as evidenced by the contribution of various phonon modes (including the localized mode) to the Huang-Rhys factor and lineshape function, depicted in the upper two panels of Fig. \ref{fig2}(a). \textit{For instance}, the dominant phonon modes contributing to the Huang-Rhys factor have phonon energies ranging from 25 to 45 meV, significantly lower than the phonon energy of the defect-induced localized mode. The contributions from multiple phonon modes are also reflected in the calculated electron-phonon coupling. As shown in the third panel of Fig. \ref{fig2}(a), the modes with higher phonon energy around 90 meV mainly contribute to the electron-phonon coupling (the computational detail of electron-phonon coupling is discussed in Supplemental Material \cite{supplement}). Consequently, the overall transition rate arises from a combination of both the electron-phonon coupling and the lineshape function, which gives rise to the capture coefficient $C_p$ in the order of $10^{-8} \text{ cm}^3 \text{ s}^{-1}$ at $\text{T} = 300 \text{ K}$, as illustrated in the lower panel of Fig. \ref{fig2}(a).

\par The results obtained using the equal-mode approximation based on final state can differ significantly, as shown in Fig. \ref{fig2}(b). First, compared to GaN bulk, neutral $\text{C}_{\text{N}}$ defect system also introduces a localized mode, but this mode has a lower phonon energy at $\hbar\omega = 55 \text{ meV}$ (Fig. S1). It is important to note that this mode differs not only in the direction of the lattice relaxation but also from the localized modes associated with $-1$ charged $\text{C}_{\text{N}}$ due to the phonon renormalization during defect relaxation. Second, this mode does not contribute to the Huang-Rhys factor, in contrast to the case of $-1$ charged $\text{C}_{\text{N}}$ where its localized phonon mode does contribute. Moreover, the calculated lineshape function using the final-state phonons is more than one order of magnitude smaller than that using the initial-state phonons. These discrepancies suggest that the phonon renormalization can be pronounced and the results can be contradictory when initial-state and final-state phonons are used. As a result, the capture coefficient $C_p$ is in the order of $10^{-10} \text{ cm}^3 \text{ s}^{-1}$ at $\text{T} = 300 \text{ K}$, more than two orders of magnitude smaller than the value derived from the equal-mode approximation based on initial state. In the following, we will investigate the underlying physical origins of these discrepancies when adopting the equal-mode approximation based on initial state and final state.

\begin{figure*}[htbp]
	\centering
	\includegraphics[width=0.75\textwidth]{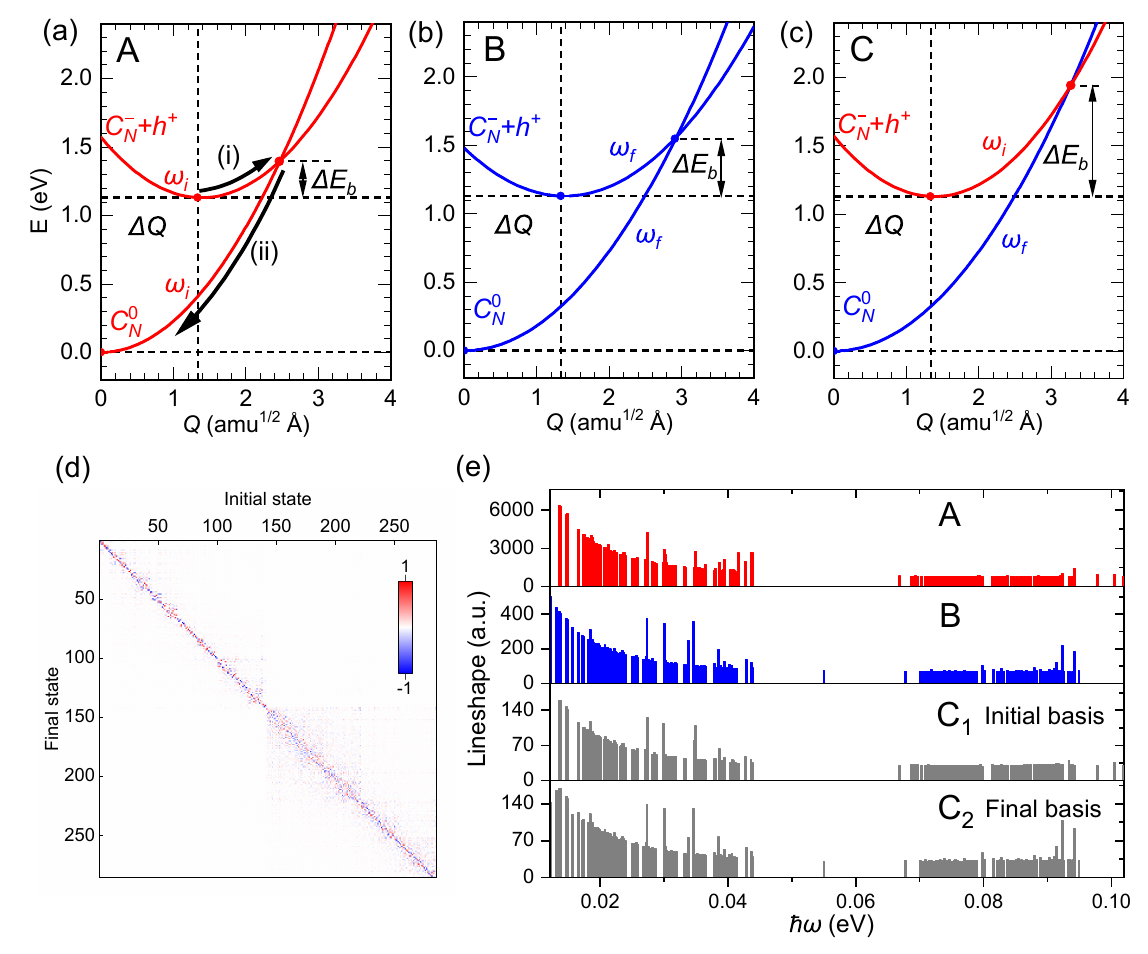}
	\caption{(a, b) Effective configuration coordinate diagrams of equal-mode approximation based on initial state and final state, and (c) that of considering mode mixing. The black arrow indicates a barrier should be overcome for multiphonon transitions in a classical picture. (d) Duschinsky rotation matrix between the phonon modes of neutral $\text{C}_{\text{N}}$ and $-1$ charged $\text{C}_{\text{N}}$. (e) Comparison of the calculated lineshape functions in different phonon modes. The ABC labels indicate the corresponding approximations or method used to calculate the lineshape function, which are consistent with the labels in (a-c).}
	\label{fig3}
\end{figure*}

\par \textit{Origin of invalidity}\textemdash Under a high-temperature limit, when the influence of electron-phonon coupling $W_{k}^{\{ if \} }$ is disregarded, the multiphonon transition rate is determined by the term $\exp(-\Delta E_b/ k_B T)$, and $\Delta E_b$ is defined as a transition barrier \cite{kubo1955application,huang1981lattice}. Here we define the transition barrier in a way described below. When phonons are calculated based on initial-state structure, the effective phonon energy during the transition is represented as the average of phonon energies in different modes, weighted by the projection to lattice relaxation,

\begin{subequations}
    \begin{align}
        \bar{\omega}_0^2=\sum_{j,\alpha ,k} \bar{\omega} _k^2 \left( \mu _{j\alpha 0}\bar{\mu} _{j\alpha k} \right) ^2 \label{omega_ave1}\\
        \mu _{j\alpha 0}=\frac{\sqrt{m_j}\Delta R_{j\alpha}}{{\sqrt{\sum_{j^\prime\alpha ^\prime} m_{j^\prime} \Delta R_{j^\prime\alpha ^\prime}^{2}}}},  \label{omega_ave2}
    \end{align}
\end{subequations}
where $\bar{\omega}_k$ is the phonon energy of initial-state mode $k$, $\bar{\mu}_{j\alpha k} $ is the eigenvector of initial-state mode $k$ projected onto the $\alpha$ direction ($\alpha$ =$x$, $y$, $z$) of the $j^{th}$ atom in the supercell, $m_j$ is the mass of the $j^{th}$ atom. $\mu_{j\alpha 0} $ is a normalized vector along lattice relaxation depicted by the structural change in Cartesian coordinate $\Delta R_{j\alpha}$. This definition of effective phonon energy has a clear physical significance, \textit{i.e.}, the contributions of phonons to a multiphonon transition are determined by how the realistic phonon mode $\mu_{j\alpha k} $ of the defect system deviates from the accepting mode $\mu_{j\alpha 0} $ associated with lattice relaxation. Following this concept, we may also define a structural displacement related to lattice relaxation, 
\begin{equation}\label{delta_Q}
    \Delta Q = \sum_{j, \alpha } \mu_{j\alpha 0} \sqrt{m_j} \Delta R_{j\alpha}.
\end{equation}
With these definitions, we can plot the configuration coordinate diagram using the equal-mode approximation (initial state), as shown in Fig. \ref{fig3}(a). Here, the effective phonon energy of the final state is essentially the same as that of the initial state, yielding a transition barrier of 0.26 eV from the intersection of the two potential energy surfaces. Similarly, we can repeat the computations in Eqs. (\ref{omega_ave1}, \ref{omega_ave2}, \ref{delta_Q}) using the final-state phonons, and obtain a transition barrier of 0.42 eV shown in Fig. \ref{fig3}(b). When we substitute these two transition barriers into the expression $\exp(-\Delta E_b/ k_B T)$, the resulting difference is approximately two orders of magnitude. This indicates that the major distinction between the initial-state and final-state equal-mode approximations stems from the inconsistent phonon overlaps reflected by the transition barriers. The comparison in such a classical picture shows the failure of equal-mode approximation, which, however, raises a significant challenge: can we achieve consistent results no matter which phonons are used?

\par \textit{Solutions}\textemdash To address the discrepancy, it is necessary to incorporate the rotation matrix (Duschinsky matrix) \cite{duschinsky1937importance,small1971herzberg} between the initial-state phonon modes and final-state phonon modes. The Duschinsky matrix has been shown to play a crucial role in phonon-assisted radiative transitions in molecules and other isolated systems \cite{small1971herzberg,baiardi2013general,borrelli2012generating,peng2007excited,liang2003influence}; however, its significance in solids, especially in semiconductors, has seldom been mentioned. Consequently, the studies on both radiative and nonradiative transitions in semiconductors have typically relied on equal-mode approximation based on initial state or final state \cite{alkauskas2012first,shi2015comparative,li2019effective}. For $\text{C}_\text{N}$ in this study, the Duschinsky matrix J between initial-state phonon mode $k$ and final-state mode $k^\prime$ can be expressed as \cite{baiardi2013general},
\begin{equation}\label{Duschinsky}
    J_{k k^\prime}=\sum_{j,\alpha}{\bar{\mu} _{j\alpha k}} \bar{\bar{\mu}} _{j\alpha k^\prime}
\end{equation}
where $\bar{\mu}_{j\alpha k}$ and $\bar{\bar{\mu}}_{j\alpha k^\prime}$ are the phonon eigenvectors of initial state and final state, respectively. In Fig. \ref{fig3}(d), we plot the Duschinsky matrix for the phonon modes in $\text{C}_\text{N}$ system. If the initial-state phonons are exactly the same as the final-state phonons, the matrix would consist solely of diagonal terms. However, the presence of non-negligible off-diagonal terms indicates that these phonons are indeed distinct, showing the necessity of incorporating them into the analysis of multiphonon transitions.

\par To incorporate the Duschinsky matrix in the classical picture, we can express the effective phonon energy of the final state based on the initial-state phonon mode,
\begin{equation}\label{omega_rotation}
    \bar{\bar{\omega}}_0^2=\sum_{j, \alpha, k^{\prime}} \bar{\bar{\omega}}_{k^{\prime}}^2\left( \mu_{j \alpha 0} \sum_k J_{k k^{\prime}} \bar{\mu}_{j \alpha k}\right)^2.
\end{equation}
With this definition of the final-state phonon energy, we plot the configuration coordinate diagram in Fig. \ref{fig3}(c). In contrast to equal-mode approximations, the inclusion of mode mixing results in a significantly larger transition barrier of 0.81 eV. The same value can be obtained when final-state phonon is set as the basis. This value differs from the barriers calculated using equal-mode approximation based on either initial state or final state. Our finding indicates that by appropriately accounting for both the initial-state and final-state phonons, a consistent result can be obtained for multiphonon transitions, regardless of which phonon is used as the basis.

\par The discussion in the classical picture above indicates that the introduction of Duschinsky matrix appears to address the discrepancy inherent in equal-mode approximation. To incorporate this effect into the transition rate calculations contributed by multiple phonon modes, it is necessary to modify the form of lattice relaxation. Specifically, to calculate the lattice relaxation matrix $\left< \chi_m(\bar{Q})\left| Q \right| \chi_n (\bar{\bar{Q}}) \right>$, the configuration $Q$ within the matrix must be projected into either the $\bar{Q}$ space or the $\bar{\bar{Q}}$ space \cite{baiardi2013general}. If we adopt the final-state phonon as the basis and then project all the phonon modes of the initial state onto those of the final state (\textit{i.e.}, performing the calculations in $\bar{\bar{Q}}$ space), the configuration in $\bar{Q}$ space can be expressed as,
\begin{equation}\label{transform}
    \bar{Q}=J\left( \bar{\bar{Q}} +\Delta \bar{\bar{Q}} \right) = J \bar{\bar{Q}} +\Delta \bar{Q},
\end{equation}
where $\Delta \bar{\bar{Q}}$ is the lattice relaxation in configuration coordinate calculated using the final-state phonon eigenvectors, \textit{i.e.}, $\Delta \bar{\bar{Q}}_k = \sum_{j,\alpha} \bar{\bar{\mu}}_{j\alpha k} \sqrt{m_j} \Delta R_{j\alpha}$, and $\Delta \bar{Q}$ is the lattice relaxation calculated using the initial-state phonon eigenvectors. Alternatively, we can also set the initial-state phonon as the basis, and project the final-state phonon modes onto the initial state, and then obtain an expression for $\bar{\bar{Q}}$ similar to Eq. (\ref{transform}). The full quantum mechanical treatment of the lattice relaxation is provided in Supplemental Material \cite{supplement}. In Fig. \ref{fig3}(e), we compare the calculated lineshape function using the equal-mode approximation based on initial state and final state, and that incorporating the Duschinsky matrix while using both the initial-state phonon and final-state phonon as the basis, respectively. The general trends in Fig. \ref{fig3}(e) are in good agreement with the transition barrier depicted in the classical picture in Fig. \ref{fig3}(a-c). For instance, the barrier calculated using equal-mode approximation (initial state) is the smallest, which correlates with the large value of lineshape function, reaching around 6000 a.u. In contrast, the equal-mode approximation (final state) gives a significantly larger barrier, corresponding to a much smaller value of lineshape function, around 400 a.u. When the Duschinsky matrix is included in the calculations, the barrier is larger than the previous two values, consistent with the lineshape function results where the maximum value for both bases reaches only about 140 a.u. regardless of the phonon basis, as shown in the lower two panels of Fig. \ref{fig3}(e).

\begin{figure}
	\centering
	\includegraphics[width=0.48\textwidth]{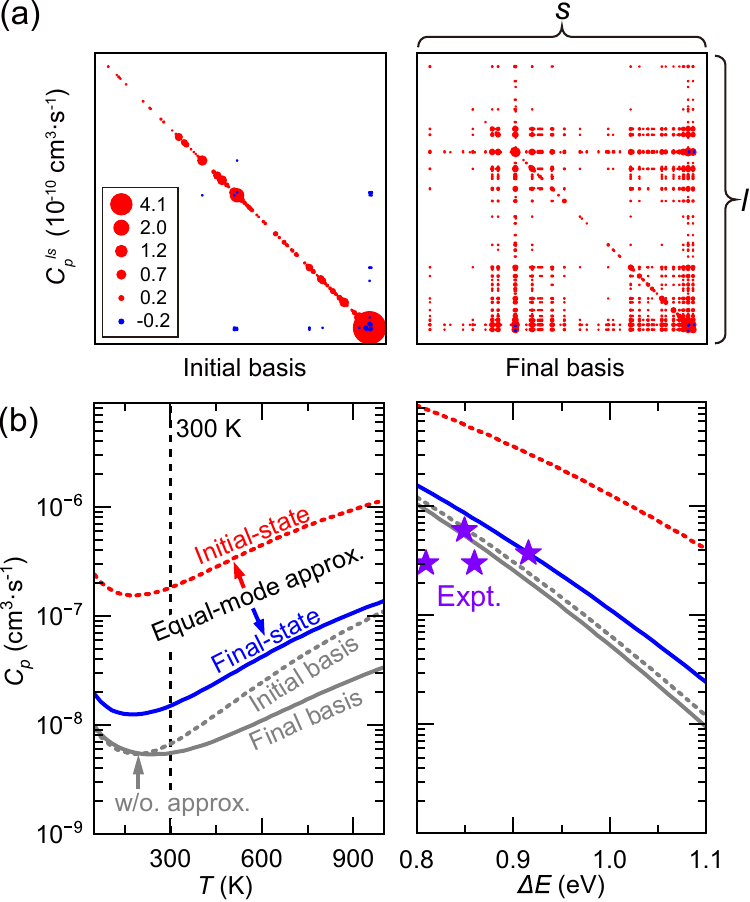}
	\caption{(a) Contribution from the off-diagonal elements $C_p^{ls}$, when the calculation is performed in the initial phonon basis (left panel) and final phonon basis (right panel). The size of the dots corresponds to the absolute value of $C_p^{ls}$, with positive values in red and negative values in blue.  (b) Hole capture coefficients as functions of temperature at $\Delta E$=1.13 eV (left panel) and as functions of transition energy at T = 300 K (right panel). The purple stars represent the experimental data \cite{reshchikov2014carrier,reshchikov2021measurement}.}
	\label{fig4}
\end{figure}

\par Besides the role of the Duschinsky matrix in the lattice relaxation, another significant factor that may reduce the error of equal-mode approximations is the off-diagonal contributions in the total transition matrix elements. Let us now focus on the lattice relaxation term $\left< \chi _m\left| Q_k \right|\chi _n \right>$. Its norm square can be written as, 
\begin{equation}
        \left| \sum_{k} \left< \chi_m\left|  Q_k \right| \chi_n \right> \right| ^2 
        = \sum_{l,s}\left< \chi _m\left| Q_l \right|\chi _n \right> \left< \chi _n\left| Q_s \right|\chi _m \right>
\end{equation}
where $l$ runs over all the initial-state phonon modes, and $s$ runs over all the final-state phonon modes. Notably, the off-diagonal elements, such as $\left< \chi _m\left| Q_1 \right|\chi _n \right> \left< \chi _n\left| Q_2 \right|\chi _m \right>$, do not vanish. Similarly, the norm square of electron-phonon coupling $W_{k}^{\{ if \} }$ also contains the off-diagonal contributions $W_{l}^{\{ if \} } W_{s}^{\{ if \} } \ (l\neq s)$. Previous studies have overlooked these cross terms. According to norm square of the total transition matrix element in Eq. (\ref{separation}), the overall transition rate depends critically on the diagonal and off-diagonal contributions from both the electron-phonon coupling term and the lattice relaxation term. In Fig. \ref{fig4}(a), We plot the matrices of hole capture coefficient $C_p^{ls}$ contributed by the initial-state phonon mode $l$ and final-state phonon mode $s$. The results indicate that off-diagonal contributions to the overall transition rate are significant and essential for achieving consistent results. The total $C_p$ is the sum of the $C_p^{ls}$ from all the diagonal and off-diagonal elements. When initial-state phonon is used as the basis during the inclusion of Duschinsky matrix, the calculated capture coefficient $C_p$ is $7.43\times10^{-9}\text{ cm}^3 \text{ s}^{-1}$ when only the diagonal elements are considered. This value slightly decreases to $7.04\times10^{-9}\text{ cm}^3 \text{ s}^{-1}$ if all the elements are considered. In contrast, when the final-state phonon is the basis, off-diagonal elements play a more significant role in overall transitions. With only diagonal elements, the capture coefficient $C_p$ is only $6.25\times10^{-10}\text{ cm}^3 \text{ s}^{-1}$, whereas considering all elements increases it by an order of magnitude to $5.52\times10^{-9}\text{ cm}^3 \text{ s}^{-1}$. This comparison shows that, with the inclusion of contributions from all off-diagonal elements in both the electron-phonon coupling and lattice relaxation matrices, the multiphonon transition rates differ by less than a factor of 2, regardless of whether the initial-state or final-state phonon is chosen as the basis.

\par In the left panel of Fig. \ref{fig4}(b), we plot the capture coefficient as a function of temperature using different methods. After incorporating the effects of the Duschinsky matrix on lattice relaxation, and accounting for the off-diagonal contributions from both the electron-phonon coupling and the lattice relaxation matrices, the capture coefficients are very close regardless of whether the initial-state or final-state phonon is used as the basis. Moreover, both values are lower than those derived from the equal-mode approximations. Given that our calculated transition energy of 1.13 eV slightly exceeds the experimentally observed range of 0.85-0.91 eV \cite{reshchikov2014carrier,reshchikov2018two,reshchikov2021measurement}, we also compute the hole capture coefficients as functions of transition energy at T = 300 K. When the transition energy falls within the experimental range, the $C_p$ calculated using our method is in good agreement with that measured in experiments.  

\par \textit{Summary}\textemdash Through systematically assessing nonradiative multiphonon processes of hole capture at $\text{C}_\text{N}$ defect in GaN, we find the phonon renormalization is significant and cannot be neglected during defect relaxation, so the widely used equal-mode approximation can cause errors by orders of magnitude in the calculated transition rate. To account for the changes of phonon modes and frequencies during transition, we introduce Duschinsky matrix between initial-state and final-state phonons and consider the off-diagonal terms of the overlap integrals. With this improvement, we demonstrate that no matter the initial-state or final-state phonon is chosen as the basis, the results are consistent with experimental values within one order of magnitude. We anticipate that this method will be widely adopted in future studies of multiphonon transitions, particularly in cases involving large defect relaxations.

\par \textit{Acknowledgments}\textemdash This work was supported by National Natural Science Foundation of China (12334005, 12188101, 12174060 and 12404089), National Key Research and Development Program of China (2022YFA1402904 and 2024YFB4205002), Science and Technology Commission of Shanghai Municipality (24JD1400600), China Postdoctoral Science Foundation Project (2023M740722), China National Postdoctoral Program for Innovative Talents (BX20230077) and Project of MOE Innovation Platform.

%\nocite{*}

\bibliography{apssamp}% Produces the bibliography via BibTeX.

%apsrev4-2.bst 2019-01-14 (MD) hand-edited version of apsrev4-1.bst
%Control: key (0)
%Control: author (72) initials jnrlst
%Control: editor formatted (1) identically to author
%Control: production of article title (-1) disabled
%Control: page (0) single
%Control: year (1) truncated
%Control: production of eprint (0) enabled
\providecommand{\noopsort}[1]{}\providecommand{\singleletter}[1]{#1}%
\begin{thebibliography}{48}%
\makeatletter
\providecommand \@ifxundefined [1]{%
 \@ifx{#1\undefined}
}%
\providecommand \@ifnum [1]{%
 \ifnum #1\expandafter \@firstoftwo
 \else \expandafter \@secondoftwo
 \fi
}%
\providecommand \@ifx [1]{%
 \ifx #1\expandafter \@firstoftwo
 \else \expandafter \@secondoftwo
 \fi
}%
\providecommand \natexlab [1]{#1}%
\providecommand \enquote  [1]{``#1''}%
\providecommand \bibnamefont  [1]{#1}%
\providecommand \bibfnamefont [1]{#1}%
\providecommand \citenamefont [1]{#1}%
\providecommand \href@noop [0]{\@secondoftwo}%
\providecommand \href [0]{\begingroup \@sanitize@url \@href}%
\providecommand \@href[1]{\@@startlink{#1}\@@href}%
\providecommand \@@href[1]{\endgroup#1\@@endlink}%
\providecommand \@sanitize@url [0]{\catcode `\\12\catcode `\$12\catcode `\&12\catcode `\#12\catcode `\^12\catcode `\_12\catcode `\%12\relax}%
\providecommand \@@startlink[1]{}%
\providecommand \@@endlink[0]{}%
\providecommand \url  [0]{\begingroup\@sanitize@url \@url }%
\providecommand \@url [1]{\endgroup\@href {#1}{\urlprefix }}%
\providecommand \urlprefix  [0]{URL }%
\providecommand \Eprint [0]{\href }%
\providecommand \doibase [0]{https://doi.org/}%
\providecommand \selectlanguage [0]{\@gobble}%
\providecommand \bibinfo  [0]{\@secondoftwo}%
\providecommand \bibfield  [0]{\@secondoftwo}%
\providecommand \translation [1]{[#1]}%
\providecommand \BibitemOpen [0]{}%
\providecommand \bibitemStop [0]{}%
\providecommand \bibitemNoStop [0]{.\EOS\space}%
\providecommand \EOS [0]{\spacefactor3000\relax}%
\providecommand \BibitemShut  [1]{\csname bibitem#1\endcsname}%
\let\auto@bib@innerbib\@empty
%</preamble>
\bibitem [{\citenamefont {Queisser}\ and\ \citenamefont {Haller}(1998)}]{defectHans1998}%
  \BibitemOpen
  \bibfield  {author} {\bibinfo {author} {\bibfnamefont {H.~J.}\ \bibnamefont {Queisser}}\ and\ \bibinfo {author} {\bibfnamefont {E.~E.}\ \bibnamefont {Haller}},\ }\href {https://doi.org/10.1126/science.281.5379.945} {\bibfield  {journal} {\bibinfo  {journal} {Science}\ }\textbf {\bibinfo {volume} {281}},\ \bibinfo {pages} {945} (\bibinfo {year} {1998})}\BibitemShut {NoStop}%
\bibitem [{\citenamefont {Grasser}(2012)}]{GRASSER201239}%
  \BibitemOpen
  \bibfield  {author} {\bibinfo {author} {\bibfnamefont {T.}~\bibnamefont {Grasser}},\ }\href {https://doi.org/https://doi.org/10.1016/j.microrel.2011.09.002} {\bibfield  {journal} {\bibinfo  {journal} {Microelectronics Reliability}\ }\textbf {\bibinfo {volume} {52}},\ \bibinfo {pages} {39} (\bibinfo {year} {2012})}\BibitemShut {NoStop}%
\bibitem [{\citenamefont {Henry}\ and\ \citenamefont {Lang}(1977)}]{henry1977nonradiative}%
  \BibitemOpen
  \bibfield  {author} {\bibinfo {author} {\bibfnamefont {C.~H.}\ \bibnamefont {Henry}}\ and\ \bibinfo {author} {\bibfnamefont {D.~V.}\ \bibnamefont {Lang}},\ }\href {https://doi.org/10.1103/PhysRevB.15.989} {\bibfield  {journal} {\bibinfo  {journal} {Phys. Rev. B}\ }\textbf {\bibinfo {volume} {15}},\ \bibinfo {pages} {989} (\bibinfo {year} {1977})}\BibitemShut {NoStop}%
\bibitem [{\citenamefont {Zhang}\ \emph {et~al.}(2021)\citenamefont {Zhang}, \citenamefont {Shen}, \citenamefont {Turiansky},\ and\ \citenamefont {Van~de Walle}}]{zhang2021minimizing}%
  \BibitemOpen
  \bibfield  {author} {\bibinfo {author} {\bibfnamefont {X.}~\bibnamefont {Zhang}}, \bibinfo {author} {\bibfnamefont {J.-X.}\ \bibnamefont {Shen}}, \bibinfo {author} {\bibfnamefont {M.~E.}\ \bibnamefont {Turiansky}},\ and\ \bibinfo {author} {\bibfnamefont {C.~G.}\ \bibnamefont {Van~de Walle}},\ }\href {https://doi.org/10.1038/s41563-021-00986-5} {\bibfield  {journal} {\bibinfo  {journal} {Nature Materials}\ }\textbf {\bibinfo {volume} {20}},\ \bibinfo {pages} {971} (\bibinfo {year} {2021})}\BibitemShut {NoStop}%
\bibitem [{\citenamefont {Demchenko}\ \emph {et~al.}(2021)\citenamefont {Demchenko}, \citenamefont {Vorobiov}, \citenamefont {Andrieiev}, \citenamefont {Myers},\ and\ \citenamefont {Reshchikov}}]{DemchenkoPRL}%
  \BibitemOpen
  \bibfield  {author} {\bibinfo {author} {\bibfnamefont {D.~O.}\ \bibnamefont {Demchenko}}, \bibinfo {author} {\bibfnamefont {M.}~\bibnamefont {Vorobiov}}, \bibinfo {author} {\bibfnamefont {O.}~\bibnamefont {Andrieiev}}, \bibinfo {author} {\bibfnamefont {T.~H.}\ \bibnamefont {Myers}},\ and\ \bibinfo {author} {\bibfnamefont {M.~A.}\ \bibnamefont {Reshchikov}},\ }\href {https://doi.org/10.1103/PhysRevLett.126.027401} {\bibfield  {journal} {\bibinfo  {journal} {Phys. Rev. Lett.}\ }\textbf {\bibinfo {volume} {126}},\ \bibinfo {pages} {027401} (\bibinfo {year} {2021})}\BibitemShut {NoStop}%
\bibitem [{\citenamefont {Huang}\ and\ \citenamefont {Rhys}(1950)}]{huang1950theory}%
  \BibitemOpen
  \bibfield  {author} {\bibinfo {author} {\bibfnamefont {K.}~\bibnamefont {Huang}}\ and\ \bibinfo {author} {\bibfnamefont {A.}~\bibnamefont {Rhys}},\ }\href {https://doi.org/10.1098/rspa.1950.0184} {\bibfield  {journal} {\bibinfo  {journal} {Proceedings of the Royal Society of London. Series A. Mathematical and Physical Sciences}\ }\textbf {\bibinfo {volume} {204}},\ \bibinfo {pages} {406} (\bibinfo {year} {1950})}\BibitemShut {NoStop}%
\bibitem [{\citenamefont {Kubo}\ and\ \citenamefont {Toyozawa}(1955)}]{kubo1955application}%
  \BibitemOpen
  \bibfield  {author} {\bibinfo {author} {\bibfnamefont {R.}~\bibnamefont {Kubo}}\ and\ \bibinfo {author} {\bibfnamefont {Y.}~\bibnamefont {Toyozawa}},\ }\href {https://doi.org/10.1143/PTP.13.160} {\bibfield  {journal} {\bibinfo  {journal} {Progress of Theoretical Physics}\ }\textbf {\bibinfo {volume} {13}},\ \bibinfo {pages} {160} (\bibinfo {year} {1955})}\BibitemShut {NoStop}%
\bibitem [{\citenamefont {P{\"a}ssler}(1974)}]{passler1974description}%
  \BibitemOpen
  \bibfield  {author} {\bibinfo {author} {\bibfnamefont {R.}~\bibnamefont {P{\"a}ssler}},\ }\href {https://doi.org/10.1007/BF01596354} {\bibfield  {journal} {\bibinfo  {journal} {Czechoslovak Journal of Physics B}\ }\textbf {\bibinfo {volume} {24}},\ \bibinfo {pages} {322} (\bibinfo {year} {1974})}\BibitemShut {NoStop}%
\bibitem [{\citenamefont {P{\"a}ssler}(1975)}]{passler1975description}%
  \BibitemOpen
  \bibfield  {author} {\bibinfo {author} {\bibfnamefont {R.}~\bibnamefont {P{\"a}ssler}},\ }\href {https://doi.org/10.1007/BF01589478} {\bibfield  {journal} {\bibinfo  {journal} {Czechoslovak Journal of Physics B}\ }\textbf {\bibinfo {volume} {25}},\ \bibinfo {pages} {219} (\bibinfo {year} {1975})}\BibitemShut {NoStop}%
\bibitem [{\citenamefont {Huang}(1981)}]{huang1981lattice}%
  \BibitemOpen
  \bibfield  {author} {\bibinfo {author} {\bibfnamefont {K.}~\bibnamefont {Huang}},\ }\href {https://doi.org/10.1080/00107518108231558} {\bibfield  {journal} {\bibinfo  {journal} {Contemporary Physics}\ }\textbf {\bibinfo {volume} {22}},\ \bibinfo {pages} {599} (\bibinfo {year} {1981})}\BibitemShut {NoStop}%
\bibitem [{\citenamefont {Razinkovas}\ \emph {et~al.}(2021)\citenamefont {Razinkovas}, \citenamefont {Doherty}, \citenamefont {Manson}, \citenamefont {Van~de Walle},\ and\ \citenamefont {Alkauskas}}]{Alkauskas2021}%
  \BibitemOpen
  \bibfield  {author} {\bibinfo {author} {\bibfnamefont {L.}~\bibnamefont {Razinkovas}}, \bibinfo {author} {\bibfnamefont {M.~W.}\ \bibnamefont {Doherty}}, \bibinfo {author} {\bibfnamefont {N.~B.}\ \bibnamefont {Manson}}, \bibinfo {author} {\bibfnamefont {C.~G.}\ \bibnamefont {Van~de Walle}},\ and\ \bibinfo {author} {\bibfnamefont {A.}~\bibnamefont {Alkauskas}},\ }\href {https://doi.org/10.1103/PhysRevB.104.045303} {\bibfield  {journal} {\bibinfo  {journal} {Phys. Rev. B}\ }\textbf {\bibinfo {volume} {104}},\ \bibinfo {pages} {045303} (\bibinfo {year} {2021})}\BibitemShut {NoStop}%
\bibitem [{\citenamefont {Shi}\ and\ \citenamefont {Wang}(2012)}]{Shi2012}%
  \BibitemOpen
  \bibfield  {author} {\bibinfo {author} {\bibfnamefont {L.}~\bibnamefont {Shi}}\ and\ \bibinfo {author} {\bibfnamefont {L.-W.}\ \bibnamefont {Wang}},\ }\href {https://doi.org/10.1103/PhysRevLett.109.245501} {\bibfield  {journal} {\bibinfo  {journal} {Phys. Rev. Lett.}\ }\textbf {\bibinfo {volume} {109}},\ \bibinfo {pages} {245501} (\bibinfo {year} {2012})}\BibitemShut {NoStop}%
\bibitem [{\citenamefont {Shi}\ \emph {et~al.}(2015)\citenamefont {Shi}, \citenamefont {Xu},\ and\ \citenamefont {Wang}}]{shi2015comparative}%
  \BibitemOpen
  \bibfield  {author} {\bibinfo {author} {\bibfnamefont {L.}~\bibnamefont {Shi}}, \bibinfo {author} {\bibfnamefont {K.}~\bibnamefont {Xu}},\ and\ \bibinfo {author} {\bibfnamefont {L.-W.}\ \bibnamefont {Wang}},\ }\href {https://doi.org/10.1103/PhysRevB.91.205315} {\bibfield  {journal} {\bibinfo  {journal} {Phys. Rev. B}\ }\textbf {\bibinfo {volume} {91}},\ \bibinfo {pages} {205315} (\bibinfo {year} {2015})}\BibitemShut {NoStop}%
\bibitem [{\citenamefont {Xiao}\ \emph {et~al.}(2020)\citenamefont {Xiao}, \citenamefont {Wang}, \citenamefont {Shi}, \citenamefont {Jiang}, \citenamefont {Li},\ and\ \citenamefont {Wang}}]{xiao2020anharmonic}%
  \BibitemOpen
  \bibfield  {author} {\bibinfo {author} {\bibfnamefont {Y.}~\bibnamefont {Xiao}}, \bibinfo {author} {\bibfnamefont {Z.}~\bibnamefont {Wang}}, \bibinfo {author} {\bibfnamefont {L.}~\bibnamefont {Shi}}, \bibinfo {author} {\bibfnamefont {X.}~\bibnamefont {Jiang}}, \bibinfo {author} {\bibfnamefont {S.}~\bibnamefont {Li}},\ and\ \bibinfo {author} {\bibfnamefont {L.}~\bibnamefont {Wang}},\ }\href@noop {} {\bibfield  {journal} {\bibinfo  {journal} {Science China Physics, Mechanics \& Astronomy}\ }\textbf {\bibinfo {volume} {63}},\ \bibinfo {pages} {277312} (\bibinfo {year} {2020})}\BibitemShut {NoStop}%
\bibitem [{\citenamefont {Qiu}\ \emph {et~al.}(2023)\citenamefont {Qiu}, \citenamefont {Song}, \citenamefont {Deng},\ and\ \citenamefont {Wei}}]{qiuchen}%
  \BibitemOpen
  \bibfield  {author} {\bibinfo {author} {\bibfnamefont {C.}~\bibnamefont {Qiu}}, \bibinfo {author} {\bibfnamefont {Y.}~\bibnamefont {Song}}, \bibinfo {author} {\bibfnamefont {H.-X.}\ \bibnamefont {Deng}},\ and\ \bibinfo {author} {\bibfnamefont {S.-H.}\ \bibnamefont {Wei}},\ }\href {https://doi.org/10.1021/jacs.3c09808} {\bibfield  {journal} {\bibinfo  {journal} {Journal of the American Chemical Society}\ }\textbf {\bibinfo {volume} {145}},\ \bibinfo {pages} {24952} (\bibinfo {year} {2023})},\ \bibinfo {note} {pMID: 37916909},\ \Eprint {https://arxiv.org/abs/https://doi.org/10.1021/jacs.3c09808} {https://doi.org/10.1021/jacs.3c09808} \BibitemShut {NoStop}%
\bibitem [{\citenamefont {Alkauskas}\ \emph {et~al.}(2014{\natexlab{a}})\citenamefont {Alkauskas}, \citenamefont {Buckley}, \citenamefont {Awschalom},\ and\ \citenamefont {de~Walle}}]{alkauskas2014NV}%
  \BibitemOpen
  \bibfield  {author} {\bibinfo {author} {\bibfnamefont {A.}~\bibnamefont {Alkauskas}}, \bibinfo {author} {\bibfnamefont {B.~B.}\ \bibnamefont {Buckley}}, \bibinfo {author} {\bibfnamefont {D.~D.}\ \bibnamefont {Awschalom}},\ and\ \bibinfo {author} {\bibfnamefont {C.~G.~V.}\ \bibnamefont {de~Walle}},\ }\href {https://doi.org/10.1088/1367-2630/16/7/073026} {\bibfield  {journal} {\bibinfo  {journal} {New Journal of Physics}\ }\textbf {\bibinfo {volume} {16}},\ \bibinfo {pages} {073026} (\bibinfo {year} {2014}{\natexlab{a}})}\BibitemShut {NoStop}%
\bibitem [{\citenamefont {Jin}\ \emph {et~al.}(2021{\natexlab{a}})\citenamefont {Jin}, \citenamefont {Govoni}, \citenamefont {Wolfowicz}, \citenamefont {Sullivan}, \citenamefont {Heremans}, \citenamefont {Awschalom},\ and\ \citenamefont {Galli}}]{Giulia2021}%
  \BibitemOpen
  \bibfield  {author} {\bibinfo {author} {\bibfnamefont {Y.}~\bibnamefont {Jin}}, \bibinfo {author} {\bibfnamefont {M.}~\bibnamefont {Govoni}}, \bibinfo {author} {\bibfnamefont {G.}~\bibnamefont {Wolfowicz}}, \bibinfo {author} {\bibfnamefont {S.~E.}\ \bibnamefont {Sullivan}}, \bibinfo {author} {\bibfnamefont {F.~J.}\ \bibnamefont {Heremans}}, \bibinfo {author} {\bibfnamefont {D.~D.}\ \bibnamefont {Awschalom}},\ and\ \bibinfo {author} {\bibfnamefont {G.}~\bibnamefont {Galli}},\ }\href {https://doi.org/10.1103/PhysRevMaterials.5.084603} {\bibfield  {journal} {\bibinfo  {journal} {Phys. Rev. Mater.}\ }\textbf {\bibinfo {volume} {5}},\ \bibinfo {pages} {084603} (\bibinfo {year} {2021}{\natexlab{a}})}\BibitemShut {NoStop}%
\bibitem [{\citenamefont {Silkinis}\ \emph {et~al.}(2024)\citenamefont {Silkinis}, \citenamefont {\ifmmode~\check{Z}\else \v{Z}\fi{}alandauskas}, \citenamefont {Thiering}, \citenamefont {Gali}, \citenamefont {Van~de Walle}, \citenamefont {Alkauskas},\ and\ \citenamefont {Razinkovas}}]{PhysRevB.110.075303}%
  \BibitemOpen
  \bibfield  {author} {\bibinfo {author} {\bibfnamefont {R.}~\bibnamefont {Silkinis}}, \bibinfo {author} {\bibfnamefont {V.}~\bibnamefont {\ifmmode~\check{Z}\else \v{Z}\fi{}alandauskas}}, \bibinfo {author} {\bibfnamefont {G.~m.~H.}\ \bibnamefont {Thiering}}, \bibinfo {author} {\bibfnamefont {A.}~\bibnamefont {Gali}}, \bibinfo {author} {\bibfnamefont {C.~G.}\ \bibnamefont {Van~de Walle}}, \bibinfo {author} {\bibfnamefont {A.}~\bibnamefont {Alkauskas}},\ and\ \bibinfo {author} {\bibfnamefont {L.}~\bibnamefont {Razinkovas}},\ }\href {https://doi.org/10.1103/PhysRevB.110.075303} {\bibfield  {journal} {\bibinfo  {journal} {Phys. Rev. B}\ }\textbf {\bibinfo {volume} {110}},\ \bibinfo {pages} {075303} (\bibinfo {year} {2024})}\BibitemShut {NoStop}%
\bibitem [{\citenamefont {Xiong}\ \emph {et~al.}(2023)\citenamefont {Xiong}, \citenamefont {Bourgois}, \citenamefont {Sheremetyeva}, \citenamefont {Chen}, \citenamefont {Dahliah}, \citenamefont {Song}, \citenamefont {Zheng}, \citenamefont {Griffin}, \citenamefont {Sipahigil},\ and\ \citenamefont {Hautier}}]{geoffory-appro}%
  \BibitemOpen
  \bibfield  {author} {\bibinfo {author} {\bibfnamefont {Y.}~\bibnamefont {Xiong}}, \bibinfo {author} {\bibfnamefont {C.}~\bibnamefont {Bourgois}}, \bibinfo {author} {\bibfnamefont {N.}~\bibnamefont {Sheremetyeva}}, \bibinfo {author} {\bibfnamefont {W.}~\bibnamefont {Chen}}, \bibinfo {author} {\bibfnamefont {D.}~\bibnamefont {Dahliah}}, \bibinfo {author} {\bibfnamefont {H.}~\bibnamefont {Song}}, \bibinfo {author} {\bibfnamefont {J.}~\bibnamefont {Zheng}}, \bibinfo {author} {\bibfnamefont {S.~M.}\ \bibnamefont {Griffin}}, \bibinfo {author} {\bibfnamefont {A.}~\bibnamefont {Sipahigil}},\ and\ \bibinfo {author} {\bibfnamefont {G.}~\bibnamefont {Hautier}},\ }\href {https://doi.org/10.1126/sciadv.adh8617} {\bibfield  {journal} {\bibinfo  {journal} {Science Advances}\ }\textbf {\bibinfo {volume} {9}},\ \bibinfo {pages} {eadh8617} (\bibinfo {year} {2023})}\BibitemShut {NoStop}%
\bibitem [{\citenamefont {MARKHAM}(1959)}]{RevModPhys.31.956}%
  \BibitemOpen
  \bibfield  {author} {\bibinfo {author} {\bibfnamefont {J.~J.}\ \bibnamefont {MARKHAM}},\ }\href {https://doi.org/10.1103/RevModPhys.31.956} {\bibfield  {journal} {\bibinfo  {journal} {Rev. Mod. Phys.}\ }\textbf {\bibinfo {volume} {31}},\ \bibinfo {pages} {956} (\bibinfo {year} {1959})}\BibitemShut {NoStop}%
\bibitem [{\citenamefont {Jin}\ \emph {et~al.}(2021{\natexlab{b}})\citenamefont {Jin}, \citenamefont {Govoni}, \citenamefont {Wolfowicz}, \citenamefont {Sullivan}, \citenamefont {Heremans}, \citenamefont {Awschalom},\ and\ \citenamefont {Galli}}]{Galli1}%
  \BibitemOpen
  \bibfield  {author} {\bibinfo {author} {\bibfnamefont {Y.}~\bibnamefont {Jin}}, \bibinfo {author} {\bibfnamefont {M.}~\bibnamefont {Govoni}}, \bibinfo {author} {\bibfnamefont {G.}~\bibnamefont {Wolfowicz}}, \bibinfo {author} {\bibfnamefont {S.~E.}\ \bibnamefont {Sullivan}}, \bibinfo {author} {\bibfnamefont {F.~J.}\ \bibnamefont {Heremans}}, \bibinfo {author} {\bibfnamefont {D.~D.}\ \bibnamefont {Awschalom}},\ and\ \bibinfo {author} {\bibfnamefont {G.}~\bibnamefont {Galli}},\ }\href {https://doi.org/10.1103/PhysRevMaterials.5.084603} {\bibfield  {journal} {\bibinfo  {journal} {Phys. Rev. Mater.}\ }\textbf {\bibinfo {volume} {5}},\ \bibinfo {pages} {084603} (\bibinfo {year} {2021}{\natexlab{b}})}\BibitemShut {NoStop}%
\bibitem [{\citenamefont {Whalley}\ \emph {et~al.}(2021)\citenamefont {Whalley}, \citenamefont {Van~Gerwen}, \citenamefont {Frost}, \citenamefont {Kim}, \citenamefont {Hood},\ and\ \citenamefont {Walsh}}]{whalley2021giant}%
  \BibitemOpen
  \bibfield  {author} {\bibinfo {author} {\bibfnamefont {L.~D.}\ \bibnamefont {Whalley}}, \bibinfo {author} {\bibfnamefont {P.}~\bibnamefont {Van~Gerwen}}, \bibinfo {author} {\bibfnamefont {J.~M.}\ \bibnamefont {Frost}}, \bibinfo {author} {\bibfnamefont {S.}~\bibnamefont {Kim}}, \bibinfo {author} {\bibfnamefont {S.~N.}\ \bibnamefont {Hood}},\ and\ \bibinfo {author} {\bibfnamefont {A.}~\bibnamefont {Walsh}},\ }\href {https://doi.org/10.1021/jacs.1c03064} {\bibfield  {journal} {\bibinfo  {journal} {Journal of the American Chemical Society}\ }\textbf {\bibinfo {volume} {143}},\ \bibinfo {pages} {9123} (\bibinfo {year} {2021})}\BibitemShut {NoStop}%
\bibitem [{\citenamefont {Zhang}\ and\ \citenamefont {Wei}(2022)}]{zhang2022origin}%
  \BibitemOpen
  \bibfield  {author} {\bibinfo {author} {\bibfnamefont {X.}~\bibnamefont {Zhang}}\ and\ \bibinfo {author} {\bibfnamefont {S.-H.}\ \bibnamefont {Wei}},\ }\href {https://doi.org/10.1103/PhysRevLett.128.136401} {\bibfield  {journal} {\bibinfo  {journal} {Phys. Rev. Lett.}\ }\textbf {\bibinfo {volume} {128}},\ \bibinfo {pages} {136401} (\bibinfo {year} {2022})}\BibitemShut {NoStop}%
\bibitem [{\citenamefont {Kresse}\ and\ \citenamefont {Furthmüller}(1996)}]{kresse1996efficiency}%
  \BibitemOpen
  \bibfield  {author} {\bibinfo {author} {\bibfnamefont {G.}~\bibnamefont {Kresse}}\ and\ \bibinfo {author} {\bibfnamefont {J.}~\bibnamefont {Furthmüller}},\ }\href {https://doi.org/https://doi.org/10.1016/0927-0256(96)00008-0} {\bibfield  {journal} {\bibinfo  {journal} {Computational Materials Science}\ }\textbf {\bibinfo {volume} {6}},\ \bibinfo {pages} {15} (\bibinfo {year} {1996})}\BibitemShut {NoStop}%
\bibitem [{\citenamefont {Kresse}\ and\ \citenamefont {Joubert}(1999)}]{kresse1999ultrasoft}%
  \BibitemOpen
  \bibfield  {author} {\bibinfo {author} {\bibfnamefont {G.}~\bibnamefont {Kresse}}\ and\ \bibinfo {author} {\bibfnamefont {D.}~\bibnamefont {Joubert}},\ }\href {https://doi.org/10.1103/PhysRevB.59.1758} {\bibfield  {journal} {\bibinfo  {journal} {Phys. Rev. B}\ }\textbf {\bibinfo {volume} {59}},\ \bibinfo {pages} {1758} (\bibinfo {year} {1999})}\BibitemShut {NoStop}%
\bibitem [{\citenamefont {Heyd}\ \emph {et~al.}(2003)\citenamefont {Heyd}, \citenamefont {Scuseria},\ and\ \citenamefont {Ernzerhof}}]{heyd2003hybrid}%
  \BibitemOpen
  \bibfield  {author} {\bibinfo {author} {\bibfnamefont {J.}~\bibnamefont {Heyd}}, \bibinfo {author} {\bibfnamefont {G.~E.}\ \bibnamefont {Scuseria}},\ and\ \bibinfo {author} {\bibfnamefont {M.}~\bibnamefont {Ernzerhof}},\ }\href {https://doi.org/10.1063/1.1564060} {\bibfield  {journal} {\bibinfo  {journal} {The Journal of Chemical Physics}\ }\textbf {\bibinfo {volume} {118}},\ \bibinfo {pages} {8207} (\bibinfo {year} {2003})}\BibitemShut {NoStop}%
\bibitem [{\citenamefont {Heyd}\ \emph {et~al.}(2006)\citenamefont {Heyd}, \citenamefont {Scuseria},\ and\ \citenamefont {Ernzerhof}}]{10.1063/1.2204597}%
  \BibitemOpen
  \bibfield  {author} {\bibinfo {author} {\bibfnamefont {J.}~\bibnamefont {Heyd}}, \bibinfo {author} {\bibfnamefont {G.~E.}\ \bibnamefont {Scuseria}},\ and\ \bibinfo {author} {\bibfnamefont {M.}~\bibnamefont {Ernzerhof}},\ }\href {https://doi.org/10.1063/1.2204597} {\bibfield  {journal} {\bibinfo  {journal} {The Journal of Chemical Physics}\ }\textbf {\bibinfo {volume} {124}},\ \bibinfo {pages} {219906} (\bibinfo {year} {2006})}\BibitemShut {NoStop}%
\bibitem [{sup()}]{supplement}%
  \BibitemOpen
  \href@noop {} {}\bibinfo {note} {See Supplemental Material at [url] for detailed and complete fomalism of multiphonon transition, and the computational treatment to electron-phonon coupling matrix. The Supplemental Material contains additional Refs. [30-34].}\BibitemShut {Stop}%
\bibitem [{\citenamefont {HUSIMI}(1940)}]{1940264}%
  \BibitemOpen
  \bibfield  {author} {\bibinfo {author} {\bibfnamefont {K.}~\bibnamefont {HUSIMI}},\ }\href {https://doi.org/10.11429/ppmsj1919.22.4_264} {\bibfield  {journal} {\bibinfo  {journal} {Nippon Sugaku-Buturigakkwai Kizi Dai 3 Ki}\ }\textbf {\bibinfo {volume} {22}},\ \bibinfo {pages} {264} (\bibinfo {year} {1940})}\BibitemShut {NoStop}%
\bibitem [{\citenamefont {Zhang}\ \emph {et~al.}(2017)\citenamefont {Zhang}, \citenamefont {Shi}, \citenamefont {Yang}, \citenamefont {Zhao}, \citenamefont {Xu},\ and\ \citenamefont {Wang}}]{zhanghaishan2017}%
  \BibitemOpen
  \bibfield  {author} {\bibinfo {author} {\bibfnamefont {H.-S.}\ \bibnamefont {Zhang}}, \bibinfo {author} {\bibfnamefont {L.}~\bibnamefont {Shi}}, \bibinfo {author} {\bibfnamefont {X.-B.}\ \bibnamefont {Yang}}, \bibinfo {author} {\bibfnamefont {Y.-J.}\ \bibnamefont {Zhao}}, \bibinfo {author} {\bibfnamefont {K.}~\bibnamefont {Xu}},\ and\ \bibinfo {author} {\bibfnamefont {L.-W.}\ \bibnamefont {Wang}},\ }\href {https://doi.org/https://doi.org/10.1002/adom.201700404} {\bibfield  {journal} {\bibinfo  {journal} {Advanced Optical Materials}\ }\textbf {\bibinfo {volume} {5}},\ \bibinfo {pages} {1700404} (\bibinfo {year} {2017})}\BibitemShut {NoStop}%
\bibitem [{\citenamefont {Freysoldt}\ \emph {et~al.}(2014)\citenamefont {Freysoldt}, \citenamefont {Grabowski}, \citenamefont {Hickel}, \citenamefont {Neugebauer}, \citenamefont {Kresse}, \citenamefont {Janotti},\ and\ \citenamefont {Van~de Walle}}]{Freysoldt2014}%
  \BibitemOpen
  \bibfield  {author} {\bibinfo {author} {\bibfnamefont {C.}~\bibnamefont {Freysoldt}}, \bibinfo {author} {\bibfnamefont {B.}~\bibnamefont {Grabowski}}, \bibinfo {author} {\bibfnamefont {T.}~\bibnamefont {Hickel}}, \bibinfo {author} {\bibfnamefont {J.}~\bibnamefont {Neugebauer}}, \bibinfo {author} {\bibfnamefont {G.}~\bibnamefont {Kresse}}, \bibinfo {author} {\bibfnamefont {A.}~\bibnamefont {Janotti}},\ and\ \bibinfo {author} {\bibfnamefont {C.~G.}\ \bibnamefont {Van~de Walle}},\ }\href {https://doi.org/10.1103/RevModPhys.86.253} {\bibfield  {journal} {\bibinfo  {journal} {Rev. Mod. Phys.}\ }\textbf {\bibinfo {volume} {86}},\ \bibinfo {pages} {253} (\bibinfo {year} {2014})}\BibitemShut {NoStop}%
\bibitem [{\citenamefont {Freysoldt}\ \emph {et~al.}(2009)\citenamefont {Freysoldt}, \citenamefont {Neugebauer},\ and\ \citenamefont {Van~de Walle}}]{FNV}%
  \BibitemOpen
  \bibfield  {author} {\bibinfo {author} {\bibfnamefont {C.}~\bibnamefont {Freysoldt}}, \bibinfo {author} {\bibfnamefont {J.}~\bibnamefont {Neugebauer}},\ and\ \bibinfo {author} {\bibfnamefont {C.~G.}\ \bibnamefont {Van~de Walle}},\ }\href {https://doi.org/10.1103/PhysRevLett.102.016402} {\bibfield  {journal} {\bibinfo  {journal} {Phys. Rev. Lett.}\ }\textbf {\bibinfo {volume} {102}},\ \bibinfo {pages} {016402} (\bibinfo {year} {2009})}\BibitemShut {NoStop}%
\bibitem [{\citenamefont {Huang}\ \emph {et~al.}(2022)\citenamefont {Huang}, \citenamefont {Zheng}, \citenamefont {Dai}, \citenamefont {Guo}, \citenamefont {Wang}, \citenamefont {Jiang}, \citenamefont {Wei},\ and\ \citenamefont {Chen}}]{Huang_2022}%
  \BibitemOpen
  \bibfield  {author} {\bibinfo {author} {\bibfnamefont {M.}~\bibnamefont {Huang}}, \bibinfo {author} {\bibfnamefont {Z.}~\bibnamefont {Zheng}}, \bibinfo {author} {\bibfnamefont {Z.}~\bibnamefont {Dai}}, \bibinfo {author} {\bibfnamefont {X.}~\bibnamefont {Guo}}, \bibinfo {author} {\bibfnamefont {S.}~\bibnamefont {Wang}}, \bibinfo {author} {\bibfnamefont {L.}~\bibnamefont {Jiang}}, \bibinfo {author} {\bibfnamefont {J.}~\bibnamefont {Wei}},\ and\ \bibinfo {author} {\bibfnamefont {S.}~\bibnamefont {Chen}},\ }\href {https://doi.org/10.1088/1674-4926/43/4/042101} {\bibfield  {journal} {\bibinfo  {journal} {Journal of Semiconductors}\ }\textbf {\bibinfo {volume} {43}},\ \bibinfo {pages} {042101} (\bibinfo {year} {2022})}\BibitemShut {NoStop}%
\bibitem [{\citenamefont {Alkauskas}\ \emph {et~al.}(2014{\natexlab{b}})\citenamefont {Alkauskas}, \citenamefont {Yan},\ and\ \citenamefont {Van~de Walle}}]{alkauskas2014first}%
  \BibitemOpen
  \bibfield  {author} {\bibinfo {author} {\bibfnamefont {A.}~\bibnamefont {Alkauskas}}, \bibinfo {author} {\bibfnamefont {Q.}~\bibnamefont {Yan}},\ and\ \bibinfo {author} {\bibfnamefont {C.~G.}\ \bibnamefont {Van~de Walle}},\ }\href {https://doi.org/10.1103/PhysRevB.90.075202} {\bibfield  {journal} {\bibinfo  {journal} {Phys. Rev. B}\ }\textbf {\bibinfo {volume} {90}},\ \bibinfo {pages} {075202} (\bibinfo {year} {2014}{\natexlab{b}})}\BibitemShut {NoStop}%
\bibitem [{\citenamefont {Turiansky}\ \emph {et~al.}(2021)\citenamefont {Turiansky}, \citenamefont {Alkauskas}, \citenamefont {Engel}, \citenamefont {Kresse}, \citenamefont {Wickramaratne}, \citenamefont {Shen}, \citenamefont {Dreyer},\ and\ \citenamefont {{Van de Walle}}}]{turiansky2021nonrad}%
  \BibitemOpen
  \bibfield  {author} {\bibinfo {author} {\bibfnamefont {M.~E.}\ \bibnamefont {Turiansky}}, \bibinfo {author} {\bibfnamefont {A.}~\bibnamefont {Alkauskas}}, \bibinfo {author} {\bibfnamefont {M.}~\bibnamefont {Engel}}, \bibinfo {author} {\bibfnamefont {G.}~\bibnamefont {Kresse}}, \bibinfo {author} {\bibfnamefont {D.}~\bibnamefont {Wickramaratne}}, \bibinfo {author} {\bibfnamefont {J.-X.}\ \bibnamefont {Shen}}, \bibinfo {author} {\bibfnamefont {C.~E.}\ \bibnamefont {Dreyer}},\ and\ \bibinfo {author} {\bibfnamefont {C.~G.}\ \bibnamefont {{Van de Walle}}},\ }\href {https://doi.org/https://doi.org/10.1016/j.cpc.2021.108056} {\bibfield  {journal} {\bibinfo  {journal} {Computer Physics Communications}\ }\textbf {\bibinfo {volume} {267}},\ \bibinfo {pages} {108056} (\bibinfo {year} {2021})}\BibitemShut {NoStop}%
\bibitem [{\citenamefont {Demchenko}\ and\ \citenamefont {Reshchikov}(2020)}]{demchenko2020koopmans}%
  \BibitemOpen
  \bibfield  {author} {\bibinfo {author} {\bibfnamefont {D.~O.}\ \bibnamefont {Demchenko}}\ and\ \bibinfo {author} {\bibfnamefont {M.~A.}\ \bibnamefont {Reshchikov}},\ }\href {https://doi.org/10.1063/1.5140661} {\bibfield  {journal} {\bibinfo  {journal} {Journal of Applied Physics}\ }\textbf {\bibinfo {volume} {127}},\ \bibinfo {pages} {155701} (\bibinfo {year} {2020})}\BibitemShut {NoStop}%
\bibitem [{\citenamefont {Reshchikov}\ \emph {et~al.}(2018)\citenamefont {Reshchikov}, \citenamefont {Vorobiov}, \citenamefont {Demchenko}, \citenamefont {\"Ozg\"ur}, \citenamefont {Morko\ifmmode~\mbox{\c{c}}\else \c{c}\fi{}}, \citenamefont {Lesnik}, \citenamefont {Hoffmann}, \citenamefont {H\"orich}, \citenamefont {Dadgar},\ and\ \citenamefont {Strittmatter}}]{reshchikov2018two}%
  \BibitemOpen
  \bibfield  {author} {\bibinfo {author} {\bibfnamefont {M.~A.}\ \bibnamefont {Reshchikov}}, \bibinfo {author} {\bibfnamefont {M.}~\bibnamefont {Vorobiov}}, \bibinfo {author} {\bibfnamefont {D.~O.}\ \bibnamefont {Demchenko}}, \bibinfo {author} {\bibfnamefont {U.}~\bibnamefont {\"Ozg\"ur}}, \bibinfo {author} {\bibfnamefont {H.}~\bibnamefont {Morko\ifmmode~\mbox{\c{c}}\else \c{c}\fi{}}}, \bibinfo {author} {\bibfnamefont {A.}~\bibnamefont {Lesnik}}, \bibinfo {author} {\bibfnamefont {M.~P.}\ \bibnamefont {Hoffmann}}, \bibinfo {author} {\bibfnamefont {F.}~\bibnamefont {H\"orich}}, \bibinfo {author} {\bibfnamefont {A.}~\bibnamefont {Dadgar}},\ and\ \bibinfo {author} {\bibfnamefont {A.}~\bibnamefont {Strittmatter}},\ }\href {https://doi.org/10.1103/PhysRevB.98.125207} {\bibfield  {journal} {\bibinfo  {journal} {Phys. Rev. B}\ }\textbf {\bibinfo {volume} {98}},\ \bibinfo {pages} {125207} (\bibinfo {year} {2018})}\BibitemShut {NoStop}%
\bibitem [{\citenamefont {Narita}\ \emph {et~al.}(2018)\citenamefont {Narita}, \citenamefont {Tomita}, \citenamefont {Tokuda}, \citenamefont {Kogiso}, \citenamefont {Horita},\ and\ \citenamefont {Kachi}}]{narita2018origin}%
  \BibitemOpen
  \bibfield  {author} {\bibinfo {author} {\bibfnamefont {T.}~\bibnamefont {Narita}}, \bibinfo {author} {\bibfnamefont {K.}~\bibnamefont {Tomita}}, \bibinfo {author} {\bibfnamefont {Y.}~\bibnamefont {Tokuda}}, \bibinfo {author} {\bibfnamefont {T.}~\bibnamefont {Kogiso}}, \bibinfo {author} {\bibfnamefont {M.}~\bibnamefont {Horita}},\ and\ \bibinfo {author} {\bibfnamefont {T.}~\bibnamefont {Kachi}},\ }\href {https://doi.org/10.1063/1.5057373} {\bibfield  {journal} {\bibinfo  {journal} {Journal of Applied Physics}\ }\textbf {\bibinfo {volume} {124}},\ \bibinfo {pages} {215701} (\bibinfo {year} {2018})}\BibitemShut {NoStop}%
\bibitem [{\citenamefont {Duschinsky}(1937)}]{duschinsky1937importance}%
  \BibitemOpen
  \bibfield  {author} {\bibinfo {author} {\bibfnamefont {F.}~\bibnamefont {Duschinsky}},\ }\href@noop {} {\bibfield  {journal} {\bibinfo  {journal} {Acta Physicochim. URSS}\ }\textbf {\bibinfo {volume} {7}},\ \bibinfo {pages} {551} (\bibinfo {year} {1937})}\BibitemShut {NoStop}%
\bibitem [{\citenamefont {Small}(1971)}]{small1971herzberg}%
  \BibitemOpen
  \bibfield  {author} {\bibinfo {author} {\bibfnamefont {G.~J.}\ \bibnamefont {Small}},\ }\href {https://doi.org/10.1063/1.1675343} {\bibfield  {journal} {\bibinfo  {journal} {The Journal of Chemical Physics}\ }\textbf {\bibinfo {volume} {54}},\ \bibinfo {pages} {3300} (\bibinfo {year} {1971})}\BibitemShut {NoStop}%
\bibitem [{\citenamefont {Baiardi}\ \emph {et~al.}(2013)\citenamefont {Baiardi}, \citenamefont {Bloino},\ and\ \citenamefont {Barone}}]{baiardi2013general}%
  \BibitemOpen
  \bibfield  {author} {\bibinfo {author} {\bibfnamefont {A.}~\bibnamefont {Baiardi}}, \bibinfo {author} {\bibfnamefont {J.}~\bibnamefont {Bloino}},\ and\ \bibinfo {author} {\bibfnamefont {V.}~\bibnamefont {Barone}},\ }\href {https://doi.org/https://doi.org/10.1016/j.cpc.2021.108056} {\bibfield  {journal} {\bibinfo  {journal} {Journal of Chemical Theory and Computation}\ }\textbf {\bibinfo {volume} {9}},\ \bibinfo {pages} {4097} (\bibinfo {year} {2013})}\BibitemShut {NoStop}%
\bibitem [{\citenamefont {Borrelli}\ \emph {et~al.}(2012)\citenamefont {Borrelli}, \citenamefont {Capobianco},\ and\ \citenamefont {Peluso}}]{borrelli2012generating}%
  \BibitemOpen
  \bibfield  {author} {\bibinfo {author} {\bibfnamefont {R.}~\bibnamefont {Borrelli}}, \bibinfo {author} {\bibfnamefont {A.}~\bibnamefont {Capobianco}},\ and\ \bibinfo {author} {\bibfnamefont {A.}~\bibnamefont {Peluso}},\ }\href {https://doi.org/doi: 10.1021/jp307887s} {\bibfield  {journal} {\bibinfo  {journal} {The Journal of Physical Chemistry A}\ }\textbf {\bibinfo {volume} {116}},\ \bibinfo {pages} {9934} (\bibinfo {year} {2012})}\BibitemShut {NoStop}%
\bibitem [{\citenamefont {Peng}\ \emph {et~al.}(2007)\citenamefont {Peng}, \citenamefont {Yi}, \citenamefont {Shuai},\ and\ \citenamefont {Shao}}]{peng2007excited}%
  \BibitemOpen
  \bibfield  {author} {\bibinfo {author} {\bibfnamefont {Q.}~\bibnamefont {Peng}}, \bibinfo {author} {\bibfnamefont {Y.}~\bibnamefont {Yi}}, \bibinfo {author} {\bibfnamefont {Z.}~\bibnamefont {Shuai}},\ and\ \bibinfo {author} {\bibfnamefont {J.}~\bibnamefont {Shao}},\ }\href {https://doi.org/10.1063/1.2710274} {\bibfield  {journal} {\bibinfo  {journal} {The Journal of Chemical Physics}\ }\textbf {\bibinfo {volume} {126}},\ \bibinfo {pages} {114302} (\bibinfo {year} {2007})}\BibitemShut {NoStop}%
\bibitem [{\citenamefont {Liang}\ \emph {et~al.}(2003)\citenamefont {Liang}, \citenamefont {Mebel}, \citenamefont {Lin}, \citenamefont {Hayashi}, \citenamefont {Selzle}, \citenamefont {Schlag},\ and\ \citenamefont {Tachiya}}]{liang2003influence}%
  \BibitemOpen
  \bibfield  {author} {\bibinfo {author} {\bibfnamefont {K.~K.}\ \bibnamefont {Liang}}, \bibinfo {author} {\bibfnamefont {A.~M.}\ \bibnamefont {Mebel}}, \bibinfo {author} {\bibfnamefont {S.~H.}\ \bibnamefont {Lin}}, \bibinfo {author} {\bibfnamefont {M.}~\bibnamefont {Hayashi}}, \bibinfo {author} {\bibfnamefont {H.~L.}\ \bibnamefont {Selzle}}, \bibinfo {author} {\bibfnamefont {E.~W.}\ \bibnamefont {Schlag}},\ and\ \bibinfo {author} {\bibfnamefont {M.}~\bibnamefont {Tachiya}},\ }\href {https://doi.org/10.1039/B305173K} {\bibfield  {journal} {\bibinfo  {journal} {Phys. Chem. Chem. Phys.}\ }\textbf {\bibinfo {volume} {5}},\ \bibinfo {pages} {4656} (\bibinfo {year} {2003})}\BibitemShut {NoStop}%
\bibitem [{\citenamefont {Alkauskas}\ \emph {et~al.}(2012)\citenamefont {Alkauskas}, \citenamefont {Lyons}, \citenamefont {Steiauf},\ and\ \citenamefont {Van~de Walle}}]{alkauskas2012first}%
  \BibitemOpen
  \bibfield  {author} {\bibinfo {author} {\bibfnamefont {A.}~\bibnamefont {Alkauskas}}, \bibinfo {author} {\bibfnamefont {J.~L.}\ \bibnamefont {Lyons}}, \bibinfo {author} {\bibfnamefont {D.}~\bibnamefont {Steiauf}},\ and\ \bibinfo {author} {\bibfnamefont {C.~G.}\ \bibnamefont {Van~de Walle}},\ }\href {https://doi.org/10.1103/PhysRevLett.109.267401} {\bibfield  {journal} {\bibinfo  {journal} {Phys. Rev. Lett.}\ }\textbf {\bibinfo {volume} {109}},\ \bibinfo {pages} {267401} (\bibinfo {year} {2012})}\BibitemShut {NoStop}%
\bibitem [{\citenamefont {Li}\ \emph {et~al.}(2019)\citenamefont {Li}, \citenamefont {Yuan}, \citenamefont {Chen}, \citenamefont {Gong},\ and\ \citenamefont {Wei}}]{li2019effective}%
  \BibitemOpen
  \bibfield  {author} {\bibinfo {author} {\bibfnamefont {J.}~\bibnamefont {Li}}, \bibinfo {author} {\bibfnamefont {Z.-K.}\ \bibnamefont {Yuan}}, \bibinfo {author} {\bibfnamefont {S.}~\bibnamefont {Chen}}, \bibinfo {author} {\bibfnamefont {X.-G.}\ \bibnamefont {Gong}},\ and\ \bibinfo {author} {\bibfnamefont {S.-H.}\ \bibnamefont {Wei}},\ }\href {https://doi.org/10.1021/acs.chemmater.8b03933} {\bibfield  {journal} {\bibinfo  {journal} {Chemistry of Materials}\ }\textbf {\bibinfo {volume} {31}},\ \bibinfo {pages} {826} (\bibinfo {year} {2019})}\BibitemShut {NoStop}%
\bibitem [{\citenamefont {Reshchikov}(2014)}]{reshchikov2014carrier}%
  \BibitemOpen
  \bibfield  {author} {\bibinfo {author} {\bibfnamefont {M.~A.}\ \bibnamefont {Reshchikov}},\ }\href {https://doi.org/10.1063/1.4865619} {\bibfield  {journal} {\bibinfo  {journal} {AIP Conference Proceedings}\ }\textbf {\bibinfo {volume} {1583}},\ \bibinfo {pages} {127} (\bibinfo {year} {2014})}\BibitemShut {NoStop}%
\bibitem [{\citenamefont {Reshchikov}(2021)}]{reshchikov2021measurement}%
  \BibitemOpen
  \bibfield  {author} {\bibinfo {author} {\bibfnamefont {M.~A.}\ \bibnamefont {Reshchikov}},\ }\href {https://doi.org/10.1063/5.0041608} {\bibfield  {journal} {\bibinfo  {journal} {Journal of Applied Physics}\ }\textbf {\bibinfo {volume} {129}},\ \bibinfo {pages} {121101} (\bibinfo {year} {2021})}\BibitemShut {NoStop}%
\end{thebibliography}%

\end{document}